\documentclass[twocolumn,showpacs]{revtex4}
\usepackage{epsfig}

\begin{document}

\title{First principles study of the origin and nature of ferromagnetism
in (Ga,Mn)As} 

\author{Stefano~Sanvito}
\email{e-mail: ssanvito@mrl.ucsb.edu} 
\affiliation{Materials Department, University of California,
Santa Barbara, CA 93106, USA$\:$}

\author{Pablo~Ordej\'on}
\affiliation{Institut de Ci\`encia de Materials de Barcelona (CSIC)
Campus de la U.A.B, E-08193 Bellaterra, Barcelona, Spain} 

\author{Nicola~A.~Hill}
\affiliation{Materials Department, University of California,
Santa Barbara, CA 93106, USA} 

\date{\today}

\begin{abstract}
The properties of diluted Ga$_{1-x}$Mn$_x$As are calculated for
a wide range of Mn concentrations within the local spin density approximation 
of density functional theory. 
M\"ulliken population analyses and orbital-resolved densities of states show 
that the configuration of Mn in GaAs is compatible with either 3d$^5$ or 3d$^6$,
however the occupation is not integer due to the large $p$-$d$ hybridization 
between the Mn $d$ states and the valence band of GaAs.  
The spin splitting of the conduction band of GaAs has a mean field-like 
linear variation with the Mn concentration and indicates ferromagnetic 
coupling with the Mn ions. 
In contrast the valence band is antiferromagnetically coupled with the Mn impurities
and the spin splitting is not linearly dependent on the Mn concentration.
This suggests that the mean field approximation breaks down in the case of 
Mn-doped GaAs and corrections due to multiple scattering must be considered. 
We calculate these corrections within a simple free electron model and find 
good agreement with our {\it ab initio} results if a large exchange constant 
($N\beta=-4.5$eV) is assumed.
\end{abstract}

\pacs{75.50.Pp, 75.30.Et, 71.15.Mb, 71.15.Fv}  
\maketitle

\section{Introduction}

The discovery of ferromagnetic order in diluted magnetic semiconductors (DMS)
made of heavily Mn-doped InAs \cite{Ohno1} and GaAs \cite{Ohno2,Ohno3,Ohno4} 
paves the way for many new semiconductor spin-devices \cite{Prin95}.
In particular the ferromagnetism of Ga$_{1-x}$Mn$_x$As adds the spin degree 
of freedom to the GaAs/(Al,Ga)As system which, in the last few years, has been 
the benchmark for new physics and for high speed electronic and optoelectronic
devices. 

Long spin-lifetime \cite{Aws1} and spin-coherence \cite{Aws2} in GaAs have 
already been demonstrated. Recently 
the feasibility of spin-injection into GaAs using 
(Ga,Mn)As contacts has been proved \cite{Aws3} overcoming the intrinsic 
difficulty of injecting spins into semiconductors from magnetic metals \cite{Schm1}. 
These two effects suggest that the GaAs/(Al,Ga)As/(Ga,Mn)As system is the best
candidate for injecting, storing and manipulating spins in entirely
solid state devices; 
a valuable step towards a practical realization of quantum computing \cite{Ddv1}.

Although there is general agreement on the carrier- (hole-) mediated origin of
the ferromagnetism in (Ga,Mn)As, the detailed mechanism is still a matter of debate 
\cite{Diet1,Diet2,Akai}. 
Recently Dietl {\it et al.} studied the ferromagnetism of III-V DMS within the 
Zener model, and obtained good agreement with existing experimental data using few 
phenomenological parameters. 
One of the key elements of the model is the mean-field Kondo-like coupling 
($p$-$d$ Hamiltonian) between the valence band of the host semiconductor and 
the magnetic impurity
\begin{equation}
H_{sp-d}=-N\beta {\vec{s}}\cdot{\vec S}\; ,
\label{eq1}
\end{equation}
where $N\beta$ is the $p$-$d$ exchange constant, $\vec{s}$ is the valence band 
electron spin and ${\vec S}$ is the impurity spin. In this model the exchange 
constant, which governs the spin splitting of the valence band of the host semiconductor,
enters quadratically in the expression for the Curie temperature. Therefore its
exact evaluation is crucial for making quantitative predictions about the
ferromagnetism in both existing and novel materials.

Unfortunately, in contrast with the case of II-VI DMS's, the experimental
determination of $N\beta$ is not conclusive, and both the sign and the magnitude
are not well known, particularly for large Mn concentrations ($x>0.01$).
From the exciton splitting in the low dilution limit ($x<0.001$) the coupling is 
found to be ferromagnetic with an exchange constant of $N\beta$=+2.5~eV \cite{Szcz1}, 
if the exchange constant for the conduction band $N\alpha$ is assumed to be +0.2~eV 
(a typical value for Mn in II-VI semiconductors). 
Reflectance magnetic circular dichroism
\cite{Ando1} and magnetoabsorption experiments \cite{Szcz2} present 
controversial results since the absorption edge splitting is strongly dependent
on the hole concentration (Moss-Burstein effect), which in turn is difficult to
determine from transport measurements because of a strong magnetoresistance up to 
very high magnetic fields \cite{Ohno3}.
Magnetotransport experiments are able to measure only the magnitude of
the exchange constant and the values obtained vary from $|N\beta|$=3.3~eV 
\cite{Mats1} to $|N\beta|$=1.5~eV \cite{Omiy1}. Finally a recent core-level 
photoemission study \cite{Oka1} of Ga$_{0.926}$Mn$_{0.074}$As gives 
$N\beta$=-1.2~eV if the Mn$^{2+}$ configuration is assumed. Nevertheless
it is worth noting that the raw data are compatible with both the 
Mn$^{2+}$ and the Mn$^{3+}$ configurations and so one cannot make a
definite determination of the sign of $N\beta$.

From a theoretical point of view, the exchange interaction for the conduction 
band results from a direct coulombic exchange and is expected to be ferromagnetic.
In contrast the exchange interaction of the valence band has a kinetic energy 
origin.
It can be described as a virtual process in which electrons from the valence
band jump onto the localized 3$d$ states of Mn \cite{Lar1}. 
Therefore the sign and strength of the coupling depend critically on the 
population of the spin-polarized Mn $d$-orbitals.
For less then half-filled $d$-orbitals Hund's rules
suggest that the coupling is ferromagnetic; for half- and more than half-filling
the coupling is antiferromagnetic. In this respect the case of Mn is rather
peculiar. Three types of Mn centers in GaAs are possible. The first two can be
seen as substitutional Mn$^{3+}$ and Mn$^{2+}$ respectively, with the former
neutral ($A^0$ with formal 3$d^4$ configuration) with spin $S=2$ and the latter 
negatively charged ($A^-$ with formal configuration 3$d^5$) with spin $S=5/2$.
The third center is obtained when $A^-$ weakly binds a hole, forming a neutral
(3$d^5+h$) complex. The $A^-$ center provides only antiferromagnetic coupling
with the valence band, while the neutral $A^0$ centers can provide either
ferromagnetic or antiferromagnetic. The ferromagnetic coupling can arise both
from the half-filled 3$d^4$ center and by hopping through the 
spin polarized bound hole in the
(3$d^5+h$) case \cite{Scz3}.

From this brief overview it is clear that a detailed description of 
the electronic structure of Mn in GaAs is crucial for understanding and 
modeling correctly the ferromagnetism of Ga$_{1-x}$Mn$_x$As.
In this paper we address this issue by calculating the ground state properties
of (Ga,Mn)As over a range of
Mn concentrations using density functional theory
(DFT) \cite{Kohn64} in the local spin density approximation (LSDA).
We use a numerical implementation of DFT based on pseudopotentials and
pseudo-atomic orbitals \cite{Siesta1,Siesta2,Siesta3}. 
Although the convergence versus basis set with localized orbitals 
is more difficult than with plane waves (where a single parameter,
the energy cutoff, determines the completeness of the basis),
the method has the great advantage of being able to handle
a large number of atoms with an accuracy comparable to plane-waves
methods. This allows
us to investigate various Mn dilutions without the need of large computer
resources. Moreover the pseudo-atomic basis is very convenient for analysis of atomic
occupation and orbital-resolved DOS, since no overlap integrals between
different bases have to be calculated.

The remainder of this 
paper is organized as follows. In the next section we provide some technical
details about the calculation method, illustrating in particular how to 
optimize the pseudo-atomic basis set. 
Then we present our results for Ga$_{1-x}$Mn$_x$As for Mn concentrations
ranging from $x=1$ to $x=0.02$. We analyze the density of states projected
onto the different orbital components, the charge distribution around the Mn
ions and we perform M\"ulliken population analyses to determine the occupation
of the $d$-orbitals of Mn. In section IV we discuss the $p$-$d$ exchange
constant and we compare our results with that expected from the Kondo-like
effective Hamiltonian (\ref{eq1}) in the mean field approximation. Then we
illustrate how the mean field
picture breaks down in the case of Mn in GaAs and how the
LDA results can be explained by a simple model which includes
multiple  scattering contributions.
Finally in section VI we present our conclusions.

\section{Computational Technique}

Since we are interested in the calculation of the electronic properties
of diluted alloys systems, we need a method that is able to handle with
sufficient accuracy a large number of atoms within a periodic supercells
approach.
To that purpose, we use a DFT approach based on pseudopotentials, and
numerical localized atomic orbitals as basis sets. This method, implemented 
in the code SIESTA \cite{Siesta1,Siesta2,Siesta3}, 
combines accuracy and a small computational cost
compared to other approaches with considerably larger computational
requirements, such as plane waves. In this approach, however, special
care must be devoted to the optimization of the basis set, in order
to obtain the desired accuracy. In this section, we describe the
optimization procedure used in this work. For all the DFT calculations
presented here, we use the Ceperley-Alder \cite{CA}
form of the exchange-correlation potential. Self-consistency is achieved
using the Pulay density mixing scheme \cite{Pul}, with a convergence 
criterion of $10^{-6}$ for the change in the elements of the density
matrix.

\subsection{Pseudopotentials}

We use the widely used
scalar relativistic Troullier-Martins pseudopotentials 
\cite{TM} with non linear core corrections \cite{Lou82} and Kleinman-Bylander
factorization \cite{KB1}. 
The reference configurations are 4$s^2$4$p^0$3$d^5$, 4$s^2$4$p^3$3$d^0$ and 
4$s^2$4$p^1$3$d^0$ respectively for Mn, As and Ga. The cutoff radii for the $s$, 
$p$ and $d$ components of the pseudopotential are: 1)Mn 2.00, 2.20 and 1.90~au,
2)As 1.90, 2.20 and 2.50~au and
3)Ga 2.10, 2.50 and 3.0~au. We check the pseudopotentials 
at the atomic level by comparing the pseudoeigenvalues with those generated by all 
electron calculations for several atomic and ionic configurations.

In order to check the transferability of the pseudopotentials just described,
we have chosen to use a plane waves method \cite{Ger1}. 
This allows us to perform essentially
converged calculations with respect to the basis set (by using a sufficiently large
energy cutoff for the plane waves). This would be difficult with the local
orbitals, since basis set convergence is much more problematic and complicated
in that case. In that way, we isolate pseudopotential effects from basis set
effects in checking the pseudopotential.  We have computed the equilibrium
lattice constant and the band structures of both GaAs and MnAs, both of them
showing good agreement with previously published results \cite{Us1} .

After testing the pseudopotential, we have also calculated the band structure for a
fixed localized orbitals basis set over a range of pseudopotentials cutoff radii.
Our results show that, as expected, the pseudopotentials that yield the best
results with plane waves also gives the best band structures with the localized
atomic orbitals.

\subsection{Basis set: Number of $\zeta$'s}

Let us now turn our attention to the pseudoatomic basis set. The procedure to
generate the numerical atomic orbitals is described in Ref. \onlinecite{Siesta4}.
The atomic orbitals are constructed as the product of an angular function
with a given angular momentum $l$ (yielding to $s$-type, $p_x, p_y, p_z$-type,
etc. orbitals), and a numerical radial function.
Several functions with the same angular and different radial form can be
considered to represent the same atomic shell, referred to as multiple-$\zeta$ 
functions.
The radial functions are determined as follows: the first $\zeta$'s are obtained
according to the scheme proposed by Sankey and Niklewsi \cite{PAO}, 
as the confined
pseudo-atomic orbitals (PAO's) which result from the DFT solution of the free
atom with the pseudopotential, and spherical potential of radius $r_c$.
The pseudo-wavefunction $\phi(r)$ constructed in such a way extends only
to distances smaller than the cutoff radius $r_c$. Note that this does not
correspond to a simple truncation, since the pseudo-wavefunction is
continuous at $r=r_c$.  The second and successive $\zeta$'s are constructing
in the split-valence spirit. They are obtained by subtracting from the
first $\zeta$ a function which reproduces the tail of the pseudoatomic
orbital for $r>r_\mathrm{DZ}$ and continues towards the origin 
as $r^l(a-br^2)$. Here $l$ is the angular momentum, $a$ and $b$
are parameters chosen to ensure the continuity, and DZ refers to 
``double-$\zeta$''. $r_\mathrm{DZ}$ is chosen in such a way that the total
norm beyond this radius has a certain value. In the present calculation
we always fix the norm beyond $r_\mathrm{DZ}$ to 15\% of the total norm, having noted
that small variations around that value do not produce any significant changes
in the total energy. Further $\zeta$'s are calculated by repeating the same
scheme. This approach is more efficient than using excited states of the
neutral atom, which can be unbound \cite{Siesta4}.

The optimization of the PAO basis is more delicate than that of its plane-wave
counterpart, where a single parameter (the energy cutoff) determines the 
accuracy and completeness of the basis. In the case of the PAO 
bases used here, several parameters determine the accuracy of
the basis: the number of $\zeta$'s for each shell, the angular
momentum components included, the confinement radii, etc. All these
must be optimized to achieve the required accuracy. As well as in
plane-waves, more complete bases produce more accurate results, but also
require larger computational resources. Since we are interested
in describing the magnetic properties of (Ga,Mn)As, it is natural
to choose a magnetic quantity as the one to monitor the convergence
of our results with respect to the basis set quality. We study the
energy difference $\Delta_\mathrm{FA}$ between the ferromagnetic and antiferromagnetic
alignments of two Mn atoms in a four atom unit cell of zincblende MnAs as a 
function of the basis set. The lattice constant is chosen to be 
$a_0 = 5.8$\AA, which is the critical lattice constant for the
half metallic behavior of MnAs \cite{Us1}.

\begin{figure}[hbtp]
\centerline{\epsfig{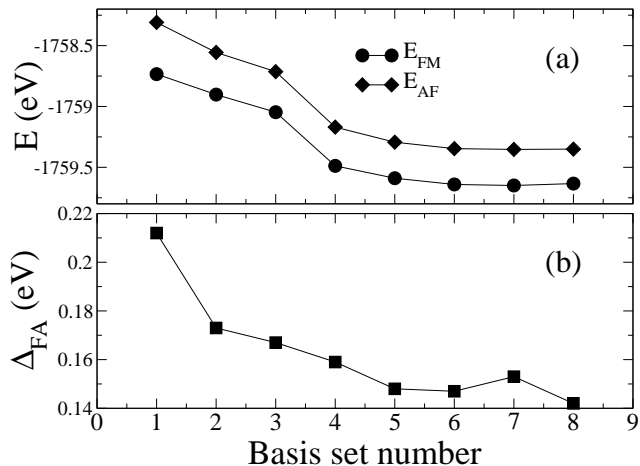}} 
\caption{\small{(a) $E_\mathrm{FM}$, $E_\mathrm{AF}$ and (b) $\Delta_\mathrm{FA}$ for the
basis of table \ref{T2}. Note the decreasing of the total energies as the
basis becomes more complete. $\Delta_\mathrm{FA}$ saturates for the basis 5.
}}
\label{F1}
\end{figure}

The first problem we address is the number of $\zeta$'s to include for each 
atomic orbital.
We start by choosing double-$\zeta$ for the $s$ orbitals of both Mn and As and
single-$\zeta$ for the $p$ orbitals and for the $d$ orbitals of Mn,
then progressively increase the number of basis orbitals. The initial
cutoff radii are chosen as indicated in table \ref{T1} and are proportional to the 
positions of the maxima of the unconstrained pseudo-wave-functions.
\begin{table}[h]
\centerline{
\begin{tabular}{c|c}
\hline
Orbital  & $r_c$ (au) \\ \hline
Mn $s$  & 6.0 \\ \hline
Mn $p$  & 6.0 \\ \hline
Mn $d$  & 5.0 \\ \hline
As $s$  & 5.5 \\ \hline
As $p$  & 5.5 \\ \hline
\end{tabular}}
\caption{\small{Initial cutoff radii used for fixing the number of $\zeta$'s. The 
radii are given in atomic unit.}}
\label{T1}
\end{table}
In figure \ref{F1} we present the total energies for the ferromagnetic 
($E_\mathrm{FM}$) and antiferromagnetic ($E_\mathrm{AF}$) alignments, and 
$\Delta_\mathrm{FA}$, for the different PAO bases listed in table \ref{T2}.
\begin{table}[hbtp]
\centerline{
\begin{tabular}{c|c|c|c|c|c}
\hline
Basis & N$_\zeta$(As $s$) & N$_\zeta$(As $p$) & N$_\zeta$(Mn $s$) &
N$_\zeta$(Mn $p$) & N$_\zeta$(Mn $d$) \\ \hline
1  & 2 & 1 & 2 & 1 & 1 \\ \hline
2  & 2 & 1 & 2 & 1 & 2  \\ \hline
3  & 2 & 1 & 2 & 2 & 2  \\ \hline
4  & 2 & 2 & 2 & 2 & 2  \\ \hline
5  & 2 & 2 & 2 & 2 & 3  \\ \hline
6  & 2 & 2 & 3 & 2 & 3  \\ \hline
7  & 2 & 3 & 2 & 2 & 3  \\ \hline
8  & 2 & 2 & 2 & 3 & 3  \\ \hline
\end{tabular}}
\caption{\small{Summary of the bases used in figure \ref{F1}. In the first column the
indicator of the basis, the following columns show the number of $\zeta$ 
(N$_\zeta$) for each orbital.}}
\label{T2}
\end{table}
From the picture two important conclusions can be reached. 
First, according to the usual variational principle the total energies for both 
ferromagnetic and antiferromagnetic configurations decrease with enlarging the 
basis.
Second the split between the ferromagnetic and antiferromagnetic configurations is 
significantly reduced by using triple-$\zeta$'s for the $d$ orbitals of Mn 
(note the large decrease of $\Delta_\mathrm{FA}$ when going from basis 1 to
basis 2 and from basis 4 to basis 5) and double-$\zeta$ for the $p$ orbitals 
of As. 
%
This sensitivity of the magnetic phase stability to the As-$p$ and Mn-$d$ basis is
consistent with the magnetism in MnAs being driven by strong $p$-$d$ hybridization. 
Since we are mainly interested in the magnetic properties of diluted systems 
describable by very large unit cells we decide to use double-$\zeta$ for all the 
orbitals except the $d$ orbitals of Mn for which we use triple-$\zeta$ 
(this is the fifth basis set in table 2).
Note that we can afford to use triple-$\zeta$ for Mn-$d$ since few Mn ions are
present in the cell. In contrast the use of larger basis sets for Ga and As yields
a more dramatic increase of the size of the computations.

\subsection{Basis set: Cutoff radii}

Next we turn our attention to the choice of the cutoff radii of the basis sets. 
For free atoms, the optimum cutoff radius of any orbital (as the one which
minimizes the energy) is infinite, since that case corresponds to no 
confinement potential, which yields to exponential tails for all the
atomic wavefunctions. However, in solids this criterion does not hold,
since the lack of vacuum and the presence of the crystal potential tend
to confine the atomic wavefunctions more than in the free atom. In this
situation, the confinement of each PAO should be optimized to minimize
the total energy. This procedure has shown in other systems like bulk bcc Fe
\cite{siesta-fe} that a finite and relatively small confinement radius
can provide lower energies and therefore more accurate bases than
long values of $r_c$. These calculations also show that the optimum confinement
radius of each PAO depend very much on the particular orbital. In our case,
however, it would be too complex to optimize the $r_c$ for all the orbitals
in our system, due to the large number of them.
Instead, we have followed a simpler approach.
We vary the cutoff radii of table \ref{T1} uniformly by multiplying all the
radii by a common scaling factor $t$. 
A somewhat similar criterion is to use as variational parameter the 
orbital energy shift $\Delta E_\mathrm{PAO}$, which is the 
energy increase that each orbital experiences when confined to a finite sphere 
and can be used as single parameter to test the convergence \cite{Siesta4}. 

\begin{figure}[hbtp]
\centerline{\epsfig{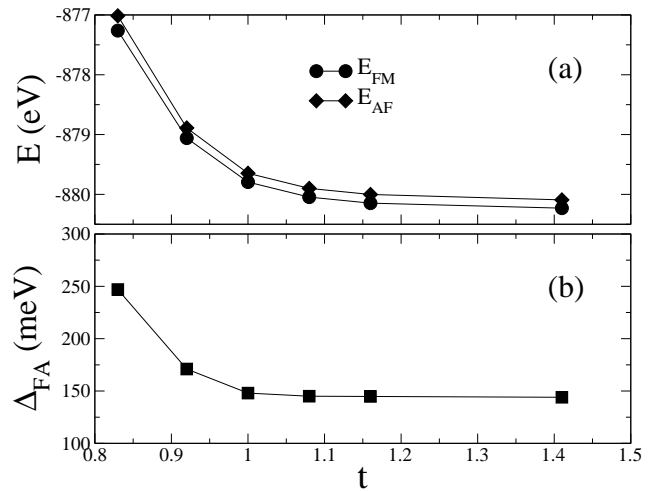}} 
\caption{\small{(a) $E_\mathrm{FM}$, $E_\mathrm{AF}$ and (b) $\Delta_\mathrm{FA}$ for the 
fifth
basis set of table \ref{T2} as a function of the scaling parameter $t$. 
Note that $\Delta_\mathrm{FA}$ saturates sooner than the total
energies for ferromagnetic and antiferromagnetic alignment.
}}
\label{F2}
\end{figure}

In figure \ref{F2} we present $E_\mathrm{FM}$, $E_\mathrm{AF}$ and 
$\Delta_\mathrm{FA}$ as a function of $t$ for the fifth basis set of table 2.
\vspace{0.0in}
The last point in figure \ref{F2} ($t=1.41$) corresponds to a basis with an 
orbital energy shift $\Delta E_\mathrm{PAO}$ of 0.001Ry. 
A convergence of 0.001Ry 
has been successfully used to describe the magnetic properties of Ni clusters 
on Ag surfaces \cite{Siesta5} and is considered an optimal value for the
convergence. However in our case we prefer to use smaller cutoff radii in order to
reduce the computation time. From figure \ref{F2} it is clear that the saturation 
of  $\Delta_{\mathrm FA}$ occurs for shorter radii than those required
to converge the total energies.
We therefore decide to fix the cutoff radii to $t=1$ noting that
$E_\mathrm{FM}$ and $E_\mathrm{AF}$ differ from the value obtained for 
$\Delta E_\mathrm{PAO}$=0.001Ry by only 0.04\% and that $\Delta_\mathrm{FA}$ differs by
only 2\%.

\subsection{Comparison with previous calculations}

We further test our basis set by computing the energy split between the Mn 
$d$ states with $e$ and $t_2$ symmetry at the $\Gamma$ point, and the 
dependence of the magnetization on the lattice spacing for zincblende MnAs. 
These two tests give an indication of the accuracy of the $p$-$d$ exchange, which 
is a dominant interaction in (Ga,Mn)As.
In fact at the $\Gamma$ point the $t_2$ states are coupled with the
As-$p$ states while the $e$ are decoupled, and their splitting is determined by the
$p$-$d$ coupling. 

\begin{figure}[hbtp]
\centerline{\epsfig{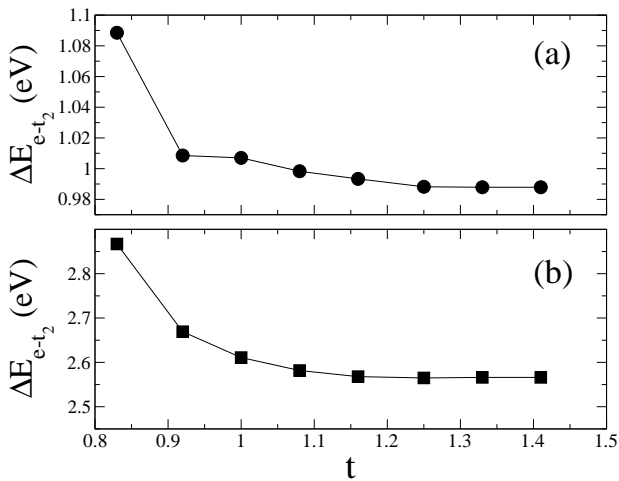}} 
\caption{\small{$e$-$t_2$ energy split at the $\Gamma$ point for MnAs: (a) majority 
spin, (b) minority spin.
}}
\label{F3}
\end{figure}

In figure \ref{F3} we present the energy split $\Delta
E_{e-t_2}=E_e-E_{t_2}$ as a function of $t$ for both the spin directions.
\vspace{0.0in}
The $e$-$t_{2}$ split converges monotonically and there is a variation of
only $\sim$2\% going from $t=1$ to $t=1.41$ ($\Delta E_\mathrm{PAO}$=0.001Ry).
If we now compare this result with our previously published results \cite{Us1}
obtained with plane-waves we notice that our present results give an $e$-$t_{2}$ 
splitting around 50meV less than the plane-wave splitting for both spins. 
This is roughly the same discrepancy 
found for $\Delta_\mathrm{FA}$.
As we have just shown, such a deviation from the plane-wave calculation cannot be 
lifted by increasing the size of the basis, since this does not produce variations 
larger than 2\%. A possible origin of such a disagreement may be the slightly 
different pseudopotentials used. 

\begin{figure}[hbtp]
\centerline{\epsfig{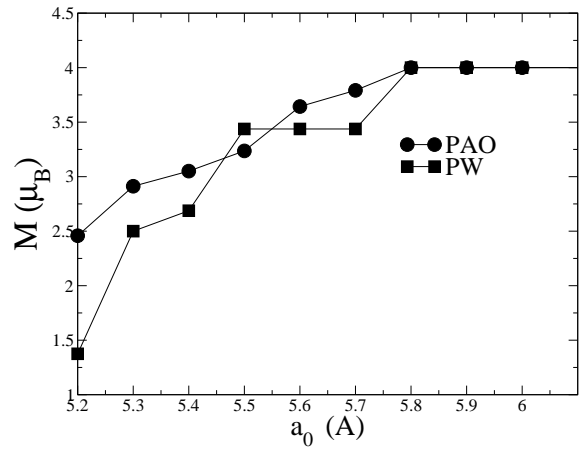}} 
\caption{\small{Magnetization as a function of the lattice constant for zincblende
MnAs.
}}
\label{F4}
\end{figure}

In figure \ref{F4} we present the magnetization as a function of the
lattice constant for MnAs and we compare it with that obtained previously
in our plane-wave calculations \cite{Us1}.
%
The agreement is quite good; the transition to the half-metallic state is 
correctly predicted for $a_0$=5.8\AA\ and also the dependence of the 
magnetization on the lattice spacing is well 
reproduced for $a_0>5.3$\AA. For smaller lattice spacings
the two calculations disagree with a tendency of the PAO basis to over-stabilize 
the ferromagnetic phase. This is a quite general behavior that we found also with
other basis sets. In particular for bases with a small number of $\zeta$'s
this effect is magnified. For instance the magnetization calculated with the
first basis set
of table \ref{T2} for $a_0=5.0$\AA\ is $\sim2.8\mu_{\mathrm B}$, while
that calculated with our optimized basis is $\sim1.4\mu_{\mathrm B}$. However
we do not believe that our plane-wave and localized basis calculations should 
necessarily agree for strongly compressed unit cells since the portability of the two 
pseudopotentials used will be different.

Finally we check the ability of our optimized basis set to describe the electronic
and structural properties of GaAs, which forms the matrix where the Mn ions
are included in (Ga,Mn)As. We find an equilibrium lattice constant of 
$a_0$=5.635\AA\ which is remarkably close to the experimental one. Moreover 
the bandstructure is very accurate; a comparison of our calculated eigenvalues
with existing calculations is presented in table
\ref{T3}.
\begin{table*}[h]
\centerline{
\begin{tabular}{c|c|c|c|c|c|c|c|c|c|c}
{\bf GaAs}  & \ $\Gamma_1$ \  & \ $\Gamma_1$ \  &  \ $X_1$ \  &  \ $X_3$ \  &  \ $X_5$ \  &  \ $X_1$ \  &  \ $L_2$ \  &
  \ $L_1$ \  &  \ $L_3$ \  &  \ $L_1$ \ \\ \hline
PAO$_\mathrm{exp}$     & -12.91 & 0.57 & -10.53 & -7.05 & -2.84 & 1.71 & -11.25 & -6.86 & -1.26 & 1.34 \ \\ \hline
PAO$_\mathrm{the}$     & -12.99 & 0.66 & -10.56 & -7.10 & -2.88 & 1.88 & -11.30 & -6.92 & -1.27 & 1.39 \ \\ \hline
LAPW \cite{sig94}  & -12.80 & 0.29 & -10.29 & -6.89 & -2.69 & 1.35 & -11.03 & -6.70 & -1.15 & 0.85 \ \\ \hline
PW-PP \cite{Fio92} & -12.56 & 0.55 & -10.25 & -6.70 & -2.58 & 1.43 & -10.95 & -6.52 & -1.09 & 1.02 \ \\ \hline
LDA-PAO \cite{S87} & -12.38 & 1.03 &  -9.85 & -6.72 & -2.66 & 1.59 & -10.63 & -6.53 & -1.14 & 1.28 \ \\ \hline
EXP  \cite{exp}    & -13.10 & 1.63 & -10.75 & -6.70 & -2.80 & 2.18 & -11.24 & -6.70 & -1.30 & 1.85 \ \\ 
\end{tabular}}
\caption{\small{Kohn-Sham eigenvalues calculated using various methods. 
The energies are calculated with respect to the top
of the valence band and all the units are eV. PAO$_\mathrm{exp}$ and PAO$_\mathrm{the}$
are the results of the present calculation assuming the lattice constant to be
respectively the experimental $a_0=5.65$\AA\ and the theoretical
$a_0=$5.635\AA.}}
\label{T3}
\end{table*}

In summary, we are confident that the results which we obtain using the
numerical atomic orbitals method with the combination of 
Troullier-Martins pseudopotentials
and the basis set described in this section, are in good agreement with
LSDA results obtained using other techniques.

\section{Electronic configuration of G\lowercase{a}A\lowercase{s}:M\lowercase{n}}

In this section we study the electronic structure of Mn in GaAs. We consider
large GaAs cells (from 2 to 96 atoms) in which we replace one Ga 
atom with a Mn atom (Mn$_\mathrm{Ga}$). 
We use 18 $k$ points in the corresponding irreducible Brillouin zones for all
the supercells and over 1000 $k$ points for the primitive zincblende unit cell
(2 atoms).
Since the cell contains only one Mn atom and we use 
periodic boundary conditions the Mn atoms are forced to be ferromagnetically aligned. 
For the smaller cells (32 and 48 atoms) we perform several simulations changing
the shape of the unit cell. This is equivalent to investigating different
arrangements of the Mn atoms with respect to each other. We find that, although
the general properties do not change, different Mn ion arrangements in the cell
result in different total energies. For all the calculations we assume the GaAs
experimental lattice spacing $a_0$=5.65\AA.

\subsection{Partial DOS and charge density distribution}

\begin{figure}[hbtp]
\centerline{\epsfig{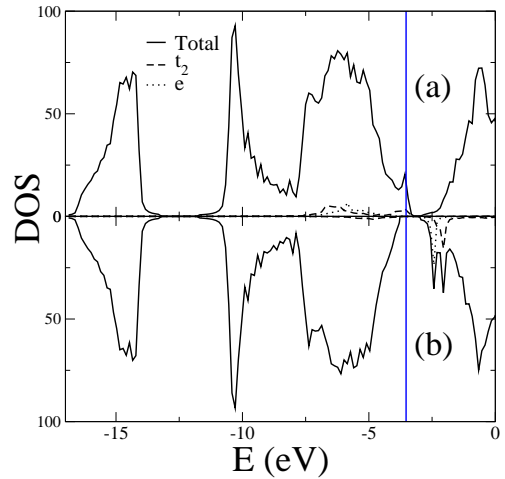}} 
\caption{\small{Partial density of state for Ga$_{1-x}$Mn$_x$As for $x=0.3$ (one 
Mn$_\mathrm{Ga}$ in a 64 atom GaAs cell): (a)
majority spin, (b) minority spin. The vertical line denotes the position of the
Fermi energy. The dashed and dotted lines represent respectively the projection
of the DOS onto the Mn-$d$ $t_2$ and $e$ orbitals.
}}
\label{F5}
\end{figure}

\begin{figure}[hbtp]
\centerline{\epsfig{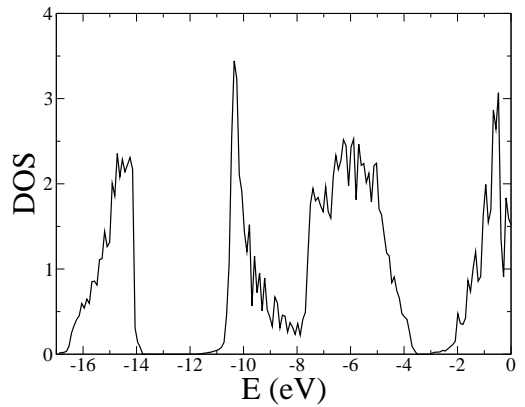}} 
\caption{\small{DOS for GaAs.
}}
\label{F5bis}
\end{figure}

We start by analyzing the orbital-resolved density of states. In
figure \ref{F5} we present as an example the DOS obtained for a 64 atom unit
cell with one Mn$_\mathrm{Ga}$ substitution. Similar features are obtained for
both higher and lower Mn concentrations.
Far from the Fermi energy the DOS remains close to the DOS of GaAs
(see figure \ref{F5bis}), with a lower
energy As-$s$ band and the Ga($s$)-As($p$) valence band. At the Fermi energy the
situation is markedly different. The majority spin band has a rather sharp 
peak, characteristic of a narrow band, while the minority spin has a gap.
Such a band structure is the signature of a half-metallic material. 
%
The total magnetization of the 
cell is 4$\mu_\mathrm{B}$. Our calculations for higher and lower Mn concentrations
show that the magnetization does not change with the Mn concentration.
In the minority band the corresponding peak is shifted to higher energy and
is very close to the edge of the GaAs conduction band.
\begin{figure}[hbtp]
\centerline{\epsfig{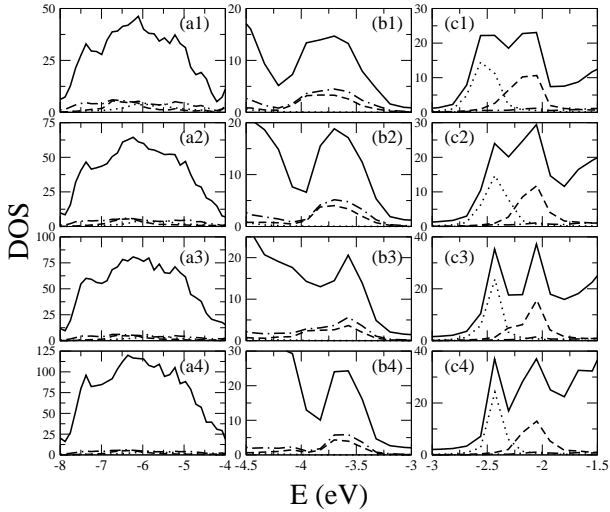}} 
\caption{\small{Total and orbital-resolved DOS for different Mn concentrations.
The three columns correspond to different energy regions and spins: 
(a) majority band between -4 and -1eV below $E_\mathrm{F}$ (broad Mn-$d$ peaks),
(b) majority band at the Fermi energy (sharp Mn-$d$ peak), (c) minority band 1eV
above $E_\mathrm{F}$. The four rows indicate different Mn concentrations: (1)
$x$=$0.06$ (1 Mn in 32 atoms), (2) $x$=$0.04$ (1 Mn in 48 atoms), (3) $x$=$0.03$
(1 Mn in 64 atom) and (4) $x$=$0.02$ (1 Mn in 96 atoms). 
The solid lines denote the total DOS while the dashed, dotted and dot-dashed lines
denote the DOS coming respectively from the $t_2$ states of Mn, the 
$e$ states of Mn, and the
As($p$) states of the four As atoms neighboring the Mn impurity.
Note that the states of columns (b) and (c) do not scale with concentration, 
indicating strong local hybridization.
}}
\label{F6}
\end{figure}
If we now consider the DOS projected onto the different orbital components of 
the wave-function and we look at the $e$ and $t_2$ $d$-states of Mn some
interesting features appear. The majority band exhibits two broad peaks
between -4eV and -1eV below the Fermi energy with strong $e$ and
$t_2$ component respectively. In addition there is a rather narrow $t_2$ peak at 
the Fermi energy. 
In contrast the minority band has almost no $d$-character below $E_\mathrm{F}$ 
but instead has two sharp $e$ and $t_2$ peaks around 1eV above $E_\mathrm{F}$.
The different peak widths reflect the different degrees of hybridization 
of the Mn-$d$ band with the GaAs bands. The hybridization is much stronger for 
states far below the Fermi energy. 

In order to have a better understanding of the $p$-$d$ hybridization in diluted
(Ga,Mn)As, we present in figure \ref{F6} the evolution of the Mn-$d$
peaks as a function of the Mn concentration.
The most relevant feature is that for the sharp peaks in both the majority and
minority bands (columns (b) and (c)) the {\it relative} intensity of the $d$ component of 
the DOS is independent of the Mn concentration.
Therefore those portions of the DOS must be derived almost entirely
from the Mn impurity and its four neighboring As atoms.
This can also be seen by looking at the DOS projected onto the $p$ states of the 
four atoms tetrahedrally coordinated with Mn (the dot-dashed line of figure
\ref{F6}). 
In summary our analysis shows that the MnAs$_4$ complex accounts for most of the
DOS at the valence band edge for the majority band and at the conduction band edge 
for the minority.

\begin{figure}[hbtp]
\centerline{\epsfig{file=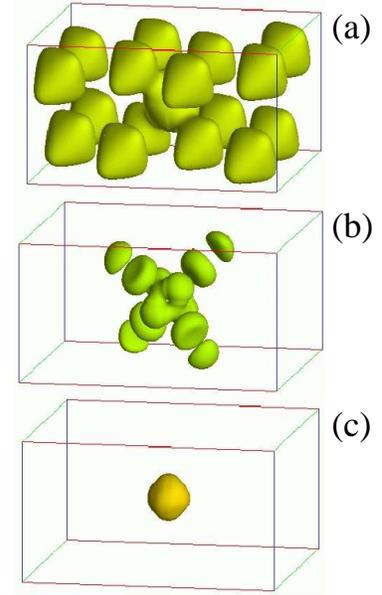,scale=0.5,angle=0}} 
\caption{\small{Charge density isosurfaces for three states shown in figure \ref{F6}. 
In this case we consider only $x$=0.06 (1 Mn in a cubic 32 atom cell). (a) Majority 
band between -4 and -1eV below $E_\mathrm{F}$ (broad Mn-$d$ peaks), (b) majority 
band at the Fermi energy (sharp Mn-$d$ peak), (c) minority band 1eV above 
$E_\mathrm{F}$. The Mn ion is in the center of the cell.}}
\label{F7}
\end{figure}

In contrast the Mn-$d$ states far below $E_\mathrm{F}$ results from strong 
coupling with the $p$ orbitals of all the As atoms of the GaAs cell.
This can be clearly seen in figure \ref{F7}, where we present the charge 
density isosurface plots corresponding to the three DOS of figure \ref{F6}.
Figure \ref{F7} shows that the charge corresponding to states at the
edge of the GaAs band gap is localized around the MnAs$_4$ complex (figures
\ref{F7}b and \ref{F7}c), while the remaining Mn-$d$ states are hybridized with 
all the As-$p$ orbitals (figure \ref{F7}a).

\begin{figure}[hbtp]
\centerline{\epsfig{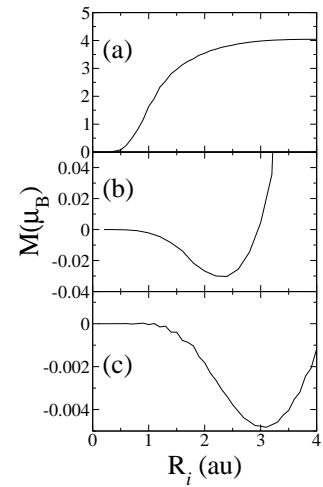}} 
\caption{\small{Magnetization profile as a function of the integration radius $R_i$ for
(Ga,Mn)As with $x$=0.3 (1 Mn ion in a cubic 64 atom GaAs cell). (a) Mn, (b)
first nearest As atom to Mn, (c) second nearest As atom to Mn.}}
\label{F8}
\end{figure}

We now turn our attention to the distribution of the magnetization around the Mn
ion. The magnetization around one atom placed at ${\bf R_0}$ is calculated as
\begin{equation}
M(R_i)=\int_{\Omega_{R_i}}[\rho_\uparrow({\bf r-R_0})
-\rho_\downarrow({\bf r-R_0})]dr \: ,
\label{eq2}
\end{equation}
where $\Omega_{R_i}$ is a sphere of radius $R_i$ and $\rho_\sigma$ is the charge
density for the spin $\sigma$. The charge density is calculated on a real-space
grid by evaluating the localized orbitals on such a grid \cite{Siesta2}. 
Of course $M(R_i)$ depends on the cutoff radius 
$R_i$. In figure \ref{F8} we present the magnetization of Mn, and of the first and
second As nearest neighbors of Mn, as a function of $R_i$.

\vspace{0.0in}

In the case of Mn the magnetization saturates for $R_i$=4.0~au and remains
almost constant up to $R_i\sim$18.0~au when the next Mn shell is encountered.
Hence we can easily deduce that the Mn magnetization is $4~\mu_\mathrm{B}$
which is the saturation value. In contrast 
the magnetization around the As ions shows a negative minimum (between 2 and 3~au 
from the As ion, depending on the position of the As ion relative
to the Mn ion) followed by a 
sharp increase. The minimum corresponds to a negative spin polarization with
respect to the Mn and the following magnetization increase occurs at
distances where the polarization of the neighboring atoms starts to be
included in the integration.
In the case of the first nearest neighbor this magnetization increase
is due mainly to the spin polarization of Mn (figure \ref{F8}b) and in the case of 
second nearest neighbors it is due to the four Ga ions coordinated to As 
(figure \ref{F8}c). 
It is interesting to note that the polarization of the As ion is always negative
with respect to that of Mn. This means that Mn and As are antiferromagnetically
coupled.
The values of the spin polarizations of As at the minima are
$-0.03~\mu_\mathrm{B}$ and $-0.005~\mu_\mathrm{B}$ respectively for first and second
nearest neighbors. These values of polarization are similar to those already
published for GaAs/MnAs superlattices calculated with a first principle LMTO-ASA
method \cite{Oga1}. It is worth noting that we did not find any sizeable changes in the 
magnetization per atom as a function of the Mn concentration for all the concentrations 
studied.

\begin{figure}[hbtp]
\centerline{\epsfig{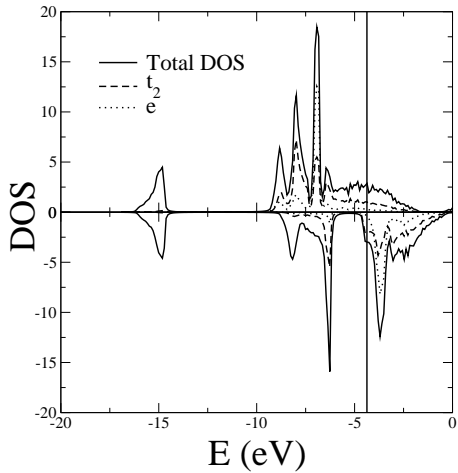}} 
\caption{\small{Orbital resolved DOS for zincblende MnAs with lattice spacing $a_0=5.65$\AA.
The vertical line denotes the position of the Fermi energy.}}
\label{F9}
\end{figure}
Finally we compare the orbital resolved DOS of (Ga,Mn)As with that of
zincblende structure
MnAs. In figure \ref{F9} we present the DOS for zincblende MnAs with the
lattice spacing of GaAs (5.65\AA), which is the same lattice spacing that we used for
diluted (Ga,Mn)As.
For this lattice spacing MnAs is not an half-metal since the Fermi energy in the 
minority band cuts through the conduction band edge, mainly dominated by $d$ 
electrons. Although the total DOS is different the projection onto the
$d$-orbitals closely resembles that of diluted (Ga,Mn)As (see figure \ref{F5}).
In particular there is a large occupation of the $d$-orbitals in the majority
band, while in the minority band only the bands of $t_2$ symmetry are occupied 
as a result of the hybridization with the As-$p$ states at the edge of the valence 
band.
The magnetization integrated around Mn ions is smaller in
zincblende MnAs at this lattice constant than in (Ga,Mn)As. 
For a lattice spacing of $a_0$=5.65~\AA\
we find a Mn polarization of $3.79~\mu_\mathrm{B}$, compared with 
$4.0~\mu_\mathrm{B}$ of (Ga,Mn)As. In contrast the polarization of As in
zincblende MnAs is considerably larger with an integrated magnetization of about 
$-0.17~\mu_\mathrm{B}$. We also note that on increasing the lattice spacing the
spin polarization of Mn increases, but the polarization of the As 
is largely unchanged. For instance at $a_0$=5.80~\AA\ we find 
$4.04~\mu_\mathrm{B}$ and $-0.17~\mu_\mathrm{B}$ respectively for the
Mn and As magnetizations.
This suggests that the polarization of Mn is related to the ionicity of the
bond with As.

A more quantitative comparison of the zincblende MnAs with diluted (Ga,Mn)As can
be obtained by performing M\"ulliken population analyses \cite{Mul1,Mul2}.
We describe the results of such analyses in the next section.

\subsection{M\"ulliken population analysis}

We perform M\"ulliken population analyses \cite{Mul1,Mul2} in order to compare
quantitatively the orbital occupations of (Ga,Mn)As at different dilutions. 
The M\"ulliken population analysis is a convenient way to separate different
contributions to the total charge density. Suppose 
we have a system described by the wave-function 
$\phi=c_1\psi_1+c_2\psi_2$,
where $\psi_\alpha$ is a localized function and $c_\alpha$ is the corresponding
amplitude. Then if the state $\phi$ is occupied by $N$ electrons, the total
occupation can be written as
\begin{equation}
N=Nc_1^2+2Nc_1c_2S_{12}+Nc_2^2\;,
\label{M2}
\end{equation}
where $S_{12}$ is the overlap integral, $\int \psi_1\psi_2dv$. M\"ulliken defined the 
sub-populations $Nc_1^2$ and $Nc_2^2$ as net populations and $2Nc_1c_2S_{12}$ as
overlap population. Moreover if the overlap population is equally splitted 
between the two wavefunctions we obtain respectively $Nc_1^2+Nc_1c_2S_{12}$ and
$Nc_2^2+Nc_1c_2S_{12}$, which are referred as gross populations. In what follows
we always refer to the gross population.
If the functions $\psi_\alpha$ represent orbital
components of the angular momentum, then the populations correspond to orbital 
populations and the overlap population is the orbital overlap population. 
Similarly if the functions $\psi_\alpha$ are the atomic wave functions for the 
atom $\alpha$ then we obtain the atomic populations and the atomic overlap 
population.
We also define M\"ulliken atomic charge as the difference between the gross
atomic charge (ie $eNc_2^2+eNc_1c_2S_{12}$ with $e$ the electronic charge) 
and the valence charge of the isolated atom.
It is widely accepted that the absolute magnitude of the atomic charges can depend
strongly on the basis set in which they are calculated \cite{Dav1}. However
relative values of M\"ulliken populations can provide useful information when
comparing different systems (for instance the amount of
covalency in semiconductors) 
\cite{Pyn1}.

We start the analysis by calculating the M\"ulliken atomic charges for Ga, Mn and As in 
GaAs, MnAs and (Ga,Mn)As at different concentrations (table \ref{T4}).  
In the case of (Ga,Mn)As for each atomic species we present the average values 
over the cell. Since we have already shown that the four As atoms coordinated
with Mn (which we denote by As$^{IV}$)
have quite different properties than the remaining As
atoms, we calculate their average atomic charge separately. 
\begin{table}[hbtp]
\centerline{
\begin{tabular}{c|c|c|c|c}
\hline
Material & Ga ($|e|$) & Mn ($|e|$) & As ($|e|$) &
As$^{IV}$ ($|e|$)  \\ \hline
GaAs                        & +0.056  &      & -0.056  &    \\ \hline
MnAs                        &   & -0.322     & +0.332  &    \\ \hline
Ga$_{0.938}$Mn$_{0.062}$As  & +0.042 & -0.089 & -0.046 & +0.005 \\ \hline
Ga$_{0.958}$Mn$_{0.042}$As  & +0.046 & -0.085 & -0.049 & +0.005 \\ \hline
Ga$_{0.969}$Mn$_{0.031}$As  & +0.047 & -0.083 & -0.049 & +0.005 \\ \hline
Ga$_{0.979}$Mn$_{0.021}$As  & +0.049 & -0.086 & -0.050 & +0.005 \\ \hline
\end{tabular}}
\caption{\small{M\"ulliken atomic charges for GaAs, MnAs and (Ga,Mn)As at different Mn
concentrations. The last two columns correspond respectively to 
the average over the As atom excluding the ones coordinated with Mn and the 
average over the four As atoms coordinated with Mn. The lattice spacing of MnAs is 
assumed 
to be $a_0$=5.65\AA.}}
\label{T4}
\end{table}
The table shows clearly the local character of the MnAs$_4$ center. 
We note that the average M\"ulliken charges of Ga and As closely resemble 
those of GaAs, particularly for low Mn concentrations. Of course in the
extremely diluted limit GaAs:Mn 
one expects the average charges of Ga and As to be exactly those
of GaAs. In contrast, the average M\"ulliken charge of the four As atoms coordinated 
with the Mn impurity does not change with concentration, confirming
that 
the electronic structure of the MnAs$_4$ complex is not affected by the 
concentration.
It is also interesting to note that these As atoms have small positive atomic 
charges whereas the other As atoms have negative atomic charge.
A positive As atomic charge is also found in zincblende
MnAs, although in that case its magnitude is much larger.
The transition from GaAs:Mn to zincblende MnAs with increasing Mn concentration
is reflected in the increase of M\"ulliken charge on the Mn atoms. Therefore the 
Mn-As bond becomes more ionic when the Mn concentration is increased. 
This picture, together with the almost complete occupation of the $d$ shells in the 
majority band discussed in the previous section, is consistent with modeling Mn in 
GaAs as an $A^0$ impurity center composed of a negatively charged Mn ion in a $d^5$ 
configuration, and a weakly bound hole ($d^5+h$) \cite{Sch1,Twa1}. 
The increase of Mn concentration, and the 
consequent increase of hole concentration, reduce the binding energy of the bound 
hole due to the partial screening of the potential. Therefore an increase of the
M\"ulliken charge of Mn with concentration is expected.
Nevertheless the agreement is only qualitative and a definitive prediction based
solely on M\"ulliken analysis cannot be made.

\begin{table*}[hbtp]
\begin{tabular}{c|c|c|c|c|c|c}
\hline
Material & Mn-$d_\uparrow$ ($|e|$) & Mn-$d_\downarrow$ ($|e|$) & As-$p_\uparrow$ ($|e|$) &
As-$p_\downarrow$ ($|e|$) & As$^{IV}$-$p_\uparrow$ ($|e|$) & As$^{IV}$-$p_\downarrow$ ($|e|$) \\ \hline
MnAs                        & 4.642 & 0.855 & 1.365 & 1.650 &  &   \\ \hline
Ga$_{0.938}$Mn$_{0.062}$As  & 4.665 & 0.788 & 1.626 & 1.637 & 1.580 & 1.647 \\ \hline
Ga$_{0.958}$Mn$_{0.042}$As  & 4.679 & 0.770 & 1.628 & 1.638 & 1.583 & 1.644 \\ \hline
Ga$_{0.969}$Mn$_{0.031}$As  & 4.675 & 0.771 & 1.630 & 1.637 & 1.584 & 1.644 \\ \hline
Ga$_{0.979}$Mn$_{0.021}$As  & 4.682 & 0.768 & 1.632 & 1.636 & 1.584 & 1.644 \\ \hline
\end{tabular}
\caption{\small{M\"ulliken atomic orbital populations in MnAs and (Ga,Mn)As at different Mn
concentrations. The symbols $\uparrow$ and $\downarrow$ correspond 
to majority and minority spin respectively.
The last two columns correspond to the four As atoms coordinated with Mn. The lattice 
spacing of MnAs is assumed to be $a_0$=5.65\AA.}}
\label{T5}
\end{table*}
We now turn our attention to the orbital population. In table \ref{T5} we present the 
orbital populations for the $p$ orbitals of As and the $d$ orbitals of Mn 
in MnAs and (Ga,Mn)As for both spin orientations. As before, we distinguish
between the As$^{IV}$ atoms and the remaining As atoms.
We do not report the orbital populations for Ga, for the $s$ orbital of As and 
for the $s$ and $p$ orbitals of Mn since they are not relevant for the 
discussion.

Several important aspects can be pointed out from the table. The total
population for the $d$ orbital of Mn is around 5.5 electronic charges for all the
systems studied. We do not expect integer values for the orbital
population since strong $p$-$d$ hybridization is present. The total overlap population
for zincblende MnAs is about 0.6 electronic charges and this can be
considered to be the uncertainty on the determination of the orbital population. This
give a Mn $d$ orbital occupation of $4.6\pm0.6$ and $0.7\pm0.6$
for the majority and minority bands respectively.
Although the orbital population is not an observable quantity and its absolute value 
may be affected by the choice of the basis set, we can conclude that the atomic 
configuration of Mn in GaAs is compatible with both 3$d^5$ and 3$d^6$. This is in
agreement with recent x-ray absorption magnetic circular dichroism
experiments \cite{Ohl1}, where the data are interpreted by assuming a Mn configuration
consisting of 80\% Mn 3$d^5$ and 20\% Mn 3$d^6$. It is interesting to note that by
decreasing the Mn concentration there is an increase of the polarization of the $d$
orbital of Mn (the orbital population is enhanced in the majority band and reduced in
the minority). This seems to be in favor of the $A^-$ 3$d^5$ configuration in 
the limit of high dilution, as is reported extensively in the literature 
\cite{Sch1,Twa1,Bha1,Sap1,Lin1}.

Table \ref{T5} also shows clearly that there is antiferromagnetic coupling between
the Mn $d$ and As $p$ orbitals. The As $p$ orbitals in fact have quite a large
spin polarization opposite to that of Mn. This cannot be due to the overlap
components of the orbital population, which would give the same polarization
as that of Mn. It is
also interesting to note that the spin polarization is much larger among the
As$^{IV}$ atoms for which it is almost insensitive to Mn
concentration, than among the other As atoms. As expected it is still smaller than
the As spin polarization in zincblende MnAs. Nevertheless we also see that
the other As atoms have a small antiferromagnetic polarization of
the $p$ orbital, which decreases with concentration as expected. This is in very good
agreement with the magnetization data presented in the previous section.

\section{Exchange Coupling}

As we pointed out in the introduction, the evaluation of the exchange constant
$N\beta$ is crucial for predicting the thermodynamic properties of (Ga,Mn)As.
In this section we provide a theoretical estimate of the exchange
constant and study its dependence on the Mn concentration. 
We begin by briefly describing the effect of the $sp$-$d$ exchange on the 
bandstructure of the host semiconductor in the mean field approximation.
Our starting point is the commonly used $sp$-$d$ exchange Hamiltonian \cite{Lar1}
\begin{eqnarray}
H_{sp-d}=-\frac{1}{2}\sum_i \sum_{n,\bf{k},\bf{k}^\prime}J_n^{sp-d}
({\bf{k}},{\bf{k}^\prime})
e^{i({{\bf k}-\bf{k}^\prime})\cdot {\bf{R}}_i}\:{\bf{S}}_i \nonumber \\
\times\left[\sum_{\mu\nu}
c_{n{\bf k}\mu}^{\dagger}\sigma_{\mu\nu}c_{n {\bf k^\prime}\nu} \right]\;,
\label{eq3}
\end{eqnarray}
where $J_n^{sp-d}({\bf k},\bf{k}^\prime)$ is the exchange integral of the band
electrons $(n,\bf{k})$ and $(n,\bf{k}^\prime)$ with the Mn local spin
${\bf{S}}_i$, $c_{n{\bf k}}^{\dagger}$ and $c_{n{\bf k}}$
are the creation and annihilation operators for an electron in band $n$ with 
Bloch vector ${\bf k}$.
The sum extends over the valence ($n=v$) and conduction 
($n=c$) bands of GaAs, and all the localized spins labelled by the index 
$i$. If we neglect interband terms which are negligible and replace the spin 
${\bf{S}}_i$ by the average spin $\langle S \rangle$ proportional to the
magnetization we restore the translational invariance of the system. 
Therefore the expression (\ref{eq3}) becomes diagonal in ${\bf{k}}$ and can be 
written as a function of the Mn fraction $x$ and the cation concentration $N$ 
as
\begin{equation}
H_{sp-d}=-\frac{1}{2}xN\langle S\rangle\sum_k J_n^{sp-d}(k)(c_{nk\uparrow}^
\dagger c_{nk\uparrow}-c_{nk\downarrow}^\dagger c_{nk\downarrow})
\;,
\label{eq4}
\end{equation}
with $\uparrow$ ($\downarrow$) indicating the up spin (down spin) direction  
with respect to the mean field spin $\langle S \rangle$ and $J_n^{sp-d}(k)=
J_n^{sp-d}(\bf{k},\bf{k})$. If we now restrict our analysis to the band-edge 
($\Gamma$ point) and define $\alpha=J_c^{sp-d}(0)$ and $\beta=J_v^{sp-d}(0)$
we obtain the equations
\begin{equation}
\begin{array}c
H_{sp-d}=-\frac{1}{2}xN\langle S\rangle \alpha(c_{c0\uparrow}^
\dagger c_{c0\uparrow}-c_{c0\downarrow}^\dagger c_{c0\downarrow})
\\
H_{sp-d}=-\frac{1}{2}xN\langle S\rangle \beta(c_{v0\uparrow}^
\dagger c_{v0\uparrow}-c_{v0\downarrow}^\dagger c_{v0\downarrow})
\end{array}
\label{eq5},
\end{equation}
for the conduction and valence bands respectively.
We note that the same analysis can be carried
out by assuming that the Mn impurities form a perfect ferromagnetic crystal. 
In such a case the derivation of equations (\ref{eq5}) is identical to that 
given here if the magnetic moment per Mn atom is used for the mean field spin
$\langle S \rangle$.
Equation (\ref{eq5}) relates the spin-splitting of the conduction and 
valence bands to the exchange integral calculated at $k=0$. This quantity
is usually extracted in optical magnetoabsorption experiments from the
spin-splitting of the exciton lines. For instance the Zeeman splitting of the
heavy hole exciton transition $E_1$ is
\begin{equation}
E_1=x\langle S\rangle N(\beta-\alpha)
\label{5b}.
\end{equation}
Other transitions give different combinations of $\alpha$ and $\beta$, which
can then be determined. Note finally that the spin-splitting of both the valence
and conduction bands in the mean field approximation is linear with the 
Mn concentration $x$.

\begin{figure}[hbtp]
\centerline{\epsfig{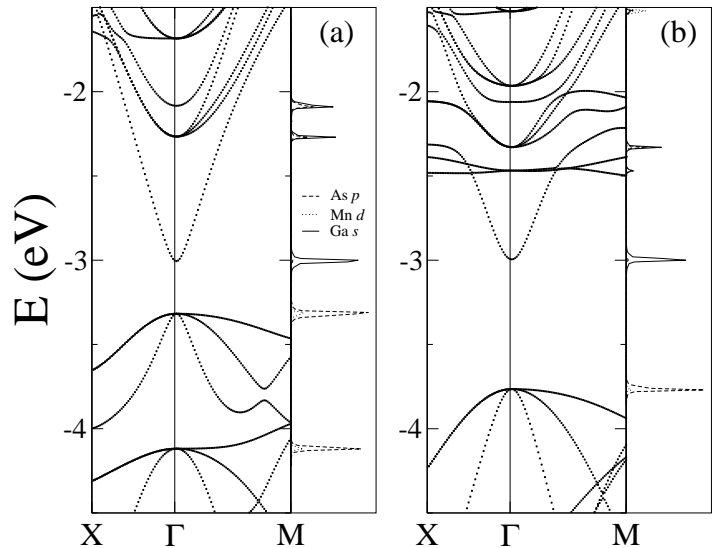}} 
\caption{\small{Band structure and orbital resolved DOS at the $\Gamma$ point
for Ga$_{1-x}$Mn$_x$As with $x$=0.3 (1 Mn ion in a cubic 64 atom GaAs cell): 
(a) majority band, (b) minority.}}
\label{F10}
\end{figure}

We calculate the exchange constants directly from the
conduction band-edge (valence band-edge) spin-splittings 
$\Delta E^c=E^c_\downarrow-E^c_\uparrow\;\;$ 
($\Delta E^v=E^v_\downarrow-E^v_\uparrow$) as follows
\begin{equation}
N\alpha=\Delta E^c/x\langle S\rangle\:, \;\;\;\;\;\;\; 
N\beta=\Delta E^v/x\langle S\rangle
\label{eq6},
\end{equation}
where $\langle S\rangle$ is half of the computed magnetization per Mn ion. 
In order to evaluate 
the parameters in equations (\ref{eq6}) we compute the band structure
around the $\Gamma$ point for large GaAs cells with a single Mn impurity. In
figure \ref{F10} we present as an example the results for a cubic cell containing
64 atoms. 
Since we are mainly interested in the $\Gamma$ point we consider 
the band structure only along the direction $(\frac{1}{8}\frac{\pi}{c_0},0,0)
\rightarrow(0,0,0)\rightarrow(\frac{1}{8}\frac{\pi}{c_0},
\frac{1}{8}\frac{\pi}{c_0},\frac{1}{8}\frac{\pi}{c_0})$ with $c_0$ the unit 
vector of the cubic cell. 
We indicate these two directions respectively as X and M. 
In figure \ref{F10} we also plot the orbital resolved
DOS at the $\Gamma$ point. This shows clearly that the valence band edge has
mainly As-$p$ character with contributions also from the $t_2$ Mn-$d$ states
due to hybridization,
while the conduction band edge is formed by Ga-$s$ states. In this way the 
spin-splitting is easily computed.

We consider different Mn concentrations and for the smaller unit cells (larger
concentrations), different geometrical arrangements. We find that the
spin-splittings of both the conduction and the valence bands are dependent
on the relative positions between the Mn ions, with variations of up to 20\%.
In particular for the same Mn concentration we find large splittings when the 
Mn ions are clustered, and smaller splittings for homogeneously
diluted systems. More details on the dependence of the exchange constant on the
spatial arrangement of the Mn ions will be published elsewhere \cite{Us3}.
In the following we consider only 
cells which maximize the separation between the Mn ions (uniform Mn
distribution). 

\begin{figure}[hbtp]
\centerline{\epsfig{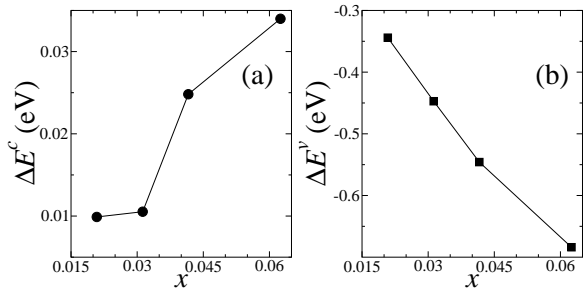}} 
\caption{\small{Spin splitting of the conduction (a) and valence (b) bands as a 
function of the Mn concentration $x$ for Ga$_{1-x}$Mn$_x$As}. The corresponding
exchange constants are listed in table \ref{T6}.}
\label{F11}
\end{figure}

In figure \ref{F11} we present the spin-splitting for the conduction and for 
the valence band as a function of the Mn concentration and in table \ref{T6}
we list the corresponding exchange constants. 
Consider first the conduction band. Although the spin splitting shows large
fluctuations with $x$, there is no systematic variation with the Mn concentration.
With the caveat that DFT is a ground state theory and therefore does not
describe accurately the conduction band, from table \ref{T6} one can conclude 
that the coupling ($s$-$d$ coupling) between the conduction band of GaAs and 
the Mn impurity is ferromagnetic. Also it is independent of $x$, as predicted
by mean field theory, and has the value $N\alpha\sim$0.2~eV.
\begin{table}[hbtp]
\centerline{
\begin{tabular}{c|c|c}
\hline
$x$       & $N\alpha$ (eV) & $N\beta$ (eV) \\ \hline
1.0       &   0.176    &  -2.44    \\ \hline
0.06250   &   0.272    &  -5.48    \\ \hline
0.04166   &   0.298    &  -6.54    \\ \hline
0.03125   &   0.168    &  -7.34    \\ \hline
0.02084   &   0.234    &  -8.16    \\ \hline
\end{tabular}}
\caption{\small{Exchange constants as a function of the Mn concentration $x$ for
Ga$_{1-x}$Mn$_x$As}.}
\label{T6}
\end{table}
Note that ferromagnetic coupling is expected since in the case of the conduction 
band the only exchange is direct and also that the value of the exchange constant 
$N\alpha$ is very close to that usually found in II-VI semiconductors
\cite{26}. 

The situation is quite different for the valence band. First of all we
see that the spin-splitting of the valence band edge is much larger than the
typical absorption edge splitting in magnetoptical experiments \cite{Szcz2}. For
instance if we compare the results for $x=0.032$ of reference \cite{Szcz2}
with those of figure \ref{F11} for the same concentration, we find that 
our calculated value is about four times larger than that obtained
experimentally. However it is
important to point out that in our calculation all the Mn ions 
contribute to the
ferromagnetism. In contrast in real systems only a fraction of the Mn
ions are ferromagnetically aligned, 
and the typical magnetization curves have a large
paramagnetic component which does not saturate even at very high magnetic fields
\cite{Oiw1}.
This has been confirmed by recent x-ray magnetic dichroism measurements
\cite{Ohl1}. 
Assuming a mean field picture, this suggests that the mean field spin calculated
here is much larger than that present in actual samples. Turning the argument
around we can conclude that our results are consistent
with experiments if we assume that
in the latter the effective Mn concentration (contributing to the ferromagnetism)
is only 1/4 of the nominal concentration.

A second important point is that, although figure
\ref{F11}b seems to suggest a linear behavior of the spin-splitting $\Delta E^c$
with $x$ according to equation (\ref{eq5}), a closer look at table \ref{T6} 
reveals that the exchange constant $N\beta$ is strongly dependent on $x$. 
Specifically, $N\beta$ increases with decreasing Mn concentration, a behavior
already well known to occur in Cd$_{1-x}$Mn$_x$S \cite{CS1,CS2,CS3}.
This dependence of the exchange constant on $x$ could be due to two
possible reasons: i)the actual $p$-$d$ coupling is not Kondo-like, 
or ii)the mean-field approximation that leads to equation
(\ref{eq4}) is not valid. 
Blinowski and Kacman studied the kinetic exchange interaction of various 3$d$
metal impurities in zincblende semiconductors \cite{Bli1}. By applying 
canonical transformations to the $p$-$d$ hybridization Hamiltonian \cite{Par1}
they evaluated the effective exchange interaction between the valence band and 
the $d$ states of the impurity. They found that for the 3$d^5$ and 3$d^6$
configurations of the impurity the effective exchange has a Kondo-like
form, while there are other non Kondo-like
contributions for the 3$d^4$ case. From
the M\"ulliken analysis we can rule out this latter configuration, and
conclude that the effective exchange is indeed Kondo-like.
Therefore the dependence of $N\beta$ on $x$ is suggestive of the breakdown of
the mean-field approximation.

\section{Breakdown of mean field approximation}

The main hypothesis sustaining the mean field approximation is that the
potential introduced by the Mn ions is weak with respect to the relevant
band-width. This seems to be true in most of the II-VI semiconductors, 
however in the case of (Cd,Mn)S such a hypothesis breaks down
and an apparent strong dependence of the exchange constant on the Mn 
concentration is found \cite{CS1,CS2,CS3}. The case of Mn in GaAs looks
very similar. We recall that for the very diluted limit there is some evidence of
the Mn ion being able to bind a polarized hole \cite{Lin1}. This suggests that 
the potential created by Mn in GaAs may be strong and hence the mean field
approximation breaks down. 

Benoit \`a la Guillaume et al. \cite{Ben1} calculated the corrections to 
the mean field approximation using a free electron model, with the 
magnetic impurities described by square potentials. The energy was calculated
within the Wigner-Seitz approach which is applicable only to the case of
perfectly periodic crystal. Although the model has been refined
\cite{Ben2,Two1} the main findings are still valid. Here we illustrate briefly
the model and we use it for computing the exchange constant.

\begin{figure}[hbtp]
\centerline{\epsfig{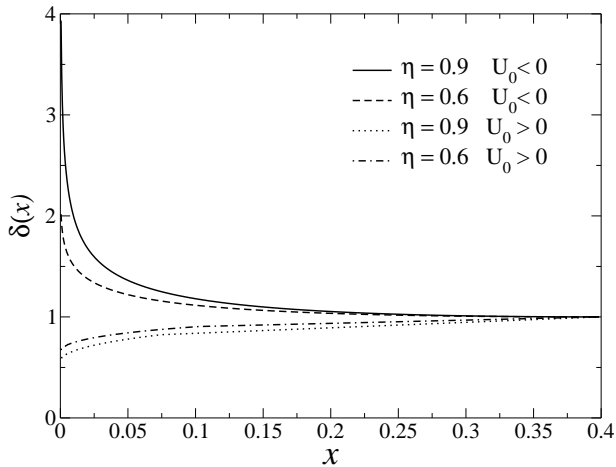}} 
\caption{\small{Dependence of the correction factor $\delta(x)$ on the
impurity concentration $x$ for a range of potentials. Note that the deviation from
mean field theory is larger for strongly attractive potentials.}}
\label{F12}
\end{figure}

We consider a free electron model with effective mass $m^*$, and uniformly
distributed magnetic impurities described by the potential
\begin{equation}
U(r)=W(r)-J(r){\bf S}\cdot{\bf s}.
\label{7}
\end{equation}
Here $W(r)$ is the spin independent substitutional potential and $J(r)$
is the $p$-$d$ coupling between the free electron spin ${\bf s}$ and the impurity
spin ${\bf S}$. We further assume that $J(r)$ and $W(r)$ have the same 
square potential shape, and that all the impurity spins are 
ferromagnetically aligned. This leads
to $U(r)=U_0\:\theta(r-b)$ and also $U^{\uparrow(\downarrow)}(r)=
(W\mp5/4J)\theta(r-b)$, when
$S=5/2$ is considered. Finally the energy is calculated by solving a
transcendental equation obtained by imposing the appropriate boundary
conditions \cite{Ben1}. We define $\delta(x,U_0)=
E(x,U_0)/E_\mathrm{mfa}(x,U_0)$ as the deviation of the computed energy 
$E(x,U_0)$ from the mean field energy $E_\mathrm{mfa}(x,U_0)=VNx$
where $V=\frac{4\pi}{3}b^3U_0$ and $Nx$ is the Mn density.
In figure \ref{F12} we present $\delta(x)$ as a
function of $x$ for different potentials $\eta=U_0/|U_c|$, where
$U_c=-(\pi \hbar/2b)^2/2m^*$ is the binding potential.  
\vspace{0.24in}
We notice that the corrections to mean field theory are large for small $x$ and
decrease with increasing $x$. In particular the mean field approximation breaks
down when the potential is attractive and close to the binding potential
($\eta\rightarrow 1$), while it is reasonably good for repulsive potentials.
We also note that the mean field approximation is recovered in both the limit of
large Mn concentrations ($x\rightarrow 1$) and in the limit of weak potential 
($\eta\rightarrow 0$). This general behavior can qualitatively explain our
LDA results. Consider in fact the band edge spin-splitting
%
\begin{eqnarray}
\Delta E^c(x)=
\frac{4\pi b^3}{3}\left[\left(W+5/4J\right)\delta(x,\eta^\downarrow)\right.-
\nonumber \\
\left.\left(W-5/4J\right)\delta(x,\eta^\uparrow)  \right]Nx\:,
\label{eq8}
\end{eqnarray}
where $\eta^{\uparrow(\downarrow)}=(W\mp5/4J)/|U_c|$. 
By comparing equation (\ref{eq8}) with figure \ref{F12} one can see that for
small $x$ the spin-splitting is largely enhanced with respect to its mean field
value. The deviation diminishes on increasing $x$, and vanishes in the limit of 
complete Mn substitution ($x=1$). Note that the application of the mean field
approximation at every $x$ gives rise to an apparent increase of the exchange
constant with the Mn concentration. This agrees with our LDA results.

It is also worth noting that the deviation from mean field theory is larger if 
the spin asymmetry of the potential $U_0$ is large. In particular the
spin-splitting is largest when the potential is attractive for one spin 
species and repulsive for the other. 
In the opposite limit, when the mean field approximation is valid 
($\delta\rightarrow 1$), the equation (\ref{eq8}) reduces to the usual expression
\begin{equation}
\Delta E^c(x)=\frac{5}{2}N\beta x\:,
\label{eq9}
\end{equation}
where we have defined the exchange constant $N\beta=N(4\pi b^3/3)J$. 

In order to
compare with experiments we perform a fit of our LDA data. We consider $b$, $W$
and $J$ as fitting parameters, with $b$ varying between the cation-anion and the
cation-cation distance and $W$ and $J$ chosen so that
no bound holes are present. This
last restriction takes into account the lack of any experimental evidence for
bound holes at the concentrations investigated here. %
\begin{figure}[hbtp]
\centerline{\epsfig{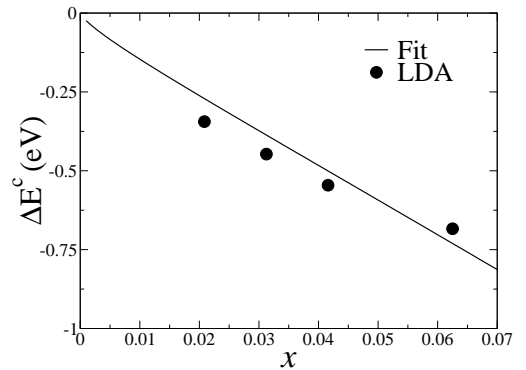}} 
\caption{\small{Band edge spin-splitting: the circles represent our LDA data and
the straight line the fit obtained with the model discussed in the text
for the parameters $b=3.6$\AA\, $J=-1.05$eV ($N\beta=-4.5$eV) and $W=-0.027$eV.}}
\label{F13}
\end{figure}
The fitting procedure yields
$b=3.6$\AA\ and $N\beta=-4.5$eV, although values in the range 3.6\AA$\:<b<3.9$\AA\ and 
$-4.9$eV$\:<N\beta<-4.4$eV fit equally well. It is important to note that for 
all the parameter
sets which give a good fit one spin hole is nearly bound while
the other feels a weak repulsive potential. Our best fit is presented in figure
\ref{F13}. Despite the roughness of the model the agreement is reasonably good. 
It is interesting to point out that the model seems to underestimate the
spin-splitting for small $x$ and overestimate it for large $x$. This is not
surprising; in the model we assume that the potential induced by the magnetic
impurity does not depend on the impurity concentration. This is in general true
for Mn in II-VI semiconductors, where Mn provides only a local spin. In the 
case of the III-V semiconductors, however,
Mn acts both as an acceptor and as source of localized spins. 
Therefore it is natural to expect a progressive screening of the Mn potential 
with concentration due to the increase of the hole density. This effect,
which is responsible for the lack of bound holes in low diluted (Ga,Mn)As, further
reduces the deviation from the mean field approximation for large $x$, and better
agreement with our LDA data may be found.

\section{Conclusions}
We have investigated theoretically the magnetic properties of Ga$_{1-x}$Mn$_x$As 
with dilutions ranging from $x$=1 to $x$=0.02. We found that Mn in GaAs has an
atomic configuration compatible with both 3$d^5$ and 3$d^6$, and that the total
occupation is not integer because of the strong $p$-$d$ coupling with the
valence band of GaAs. Such a coupling is antiferromagnetic with a
remarkably large exchange constant. We have shown that the exchange constant
has an apparent dependence on the Mn concentration. This suggests that the generally
used mean field approximation breaks down, since the potential induced by the Mn
ions in GaAs cannot be treated perturbatively. Using a simple free-electron model
we have calculated the corrections to the mean field expression for the spin-splitting of
the GaAs valence band and found a good agreement with the LDA calculations. Further
study is needed to determine the dependence of the spin-splitting on the
confinement of the Mn ions in the case of highly ordered alloys.

\section*{Acknowledgments}
This work made use of MRL Central Facilities supported by the National Science 
Foundation under award No. DMR96-32716.
This work is supported by the DARPA/ONR under the grant N0014-99-1-1096,
by ONR grant N00014-00-10557, by NSF-DMR under the grant 9973076
and by ACS PRF under the grant 33851-G5.
P. O. acknowledges support from Fundaci\'on Ram\'on Areces (Spain).
Useful discussions with N.~Samarth, D.D.~Awschalom and L.J.~Sham are kindly 
acknowledged.

\end{document}